\documentclass[%
 reprint, superscriptaddress,
 amsmath,amssymb,
 aps,
]{revtex4-2}
\pdfoutput=1 

\usepackage{graphicx}
\usepackage[dvipsnames]{xcolor}
\usepackage{subcaption}
\usepackage{dcolumn}
\usepackage{bm}
\usepackage{bbold}

\usepackage{physics}
\usepackage{amsmath,amsthm,amssymb,amsfonts,physics,siunitx,bbm,graphicx}
\usepackage{xfrac, hyperref, siunitx}
\hypersetup{
    colorlinks,
    citecolor=blue,
    filecolor=blue,
    linkcolor=blue,
    urlcolor=blue
}
\usepackage{natbib}
\usepackage{mathtools}
\usepackage{comment}

\newtheorem*{result}{Result}
\newcommand{\ba}{\begin{eqnarray}}
\newcommand{\ea}{\end{eqnarray}}
\newcommand{\ban}{\begin{eqnarray*}}
\newcommand{\ean}{\end{eqnarray*}}
\DeclareMathOperator{\pos}{pos}
\DeclareMathOperator{\sgn}{sgn}

\allowdisplaybreaks

\makeatletter
\def\maketitle{
\@author@finish
\title@column\titleblock@produce
\suppressfloats[t]}
\makeatother

\begin{document}

\title{Dynamics-Based Entanglement Witnesses for\texorpdfstring{\\}{ }Non-Gaussian States of Harmonic Oscillators}

\author{Pooja Jayachandran}
\affiliation{Centre for Quantum Technologies, National University of Singapore, 3 Science Drive 2, Singapore 117543}

\author{Lin Htoo Zaw}
\affiliation{Centre for Quantum Technologies, National University of Singapore, 3 Science Drive 2, Singapore 117543}


\author{Valerio Scarani}
\affiliation{Centre for Quantum Technologies, National University of Singapore, 3 Science Drive 2, Singapore 117543}
\affiliation{Department of Physics, National University of Singapore, 2 Science Drive 3, Singapore 117542} 

\date{\today}

\begin{abstract}
We introduce a family of entanglement witnesses for continuous variable systems, which rely on the sole assumption that their dynamics is that of coupled harmonic oscillators at the time of the test. Entanglement is inferred from the Tsirelson nonclassicality test on one of the normal modes, without any knowledge about the state of the other mode. In each round, the protocol requires measuring only the sign of one coordinate (e.g.,~position) at one among several times. This dynamic-based entanglement witness is more akin to a Bell inequality than to an uncertainty relation: in particular, it does not admit false positives from classical theory. Our criterion detects non-Gaussian states, some of which are missed by other criteria. 
\end{abstract}

\phantomsection\addcontentsline{toc}{part}{Dynamics-Based Entanglement Witnesses for Non-Gaussian States of Harmonic Oscillators}
\maketitle
\section{Introduction}\label{intro}
Once troublesome to the founders of quantum mechanics \cite{EPR1935,Schrod1935}, entanglement is now well established as one of the defining features of quantum theory. While entanglement in discrete systems has gone through much scrutiny \cite{HorodeckiEntgl,Plenio2014}, the field of continuous variable (CV) quantum entanglement has had its challenges outside the Gaussian regime \cite{Adesso_2007}.

Early entanglement criteria for CV systems were based on the second-order moments of quadrature distributions \cite{SimonGaussian,duan2000inseparability}, which are useful for Gaussian states, but ineffective otherwise \cite{GaussianQI}. However, a computation consisting only of Gaussian states and operations can be simulated efficiently on a classical computer \cite{Positive-Wigner-Simulatable}. Similar no-go theorems also exist for error correction \cite{CV-Error-Correction} and entanglement distillation \cite{CV-Entanglement-Distillation-1,CV-Entanglement-Distillation-2}, encouraging the development of methods to detect non-Gaussian CV entanglement. Efforts in this direction include criteria based on higher-order moments \cite{ZhangEtAl,hillery2006entanglement,SU2entanglement, UncertaintyIsPPT1,UncertaintyIsPPT2} and full probability distribution \cite{EntropicEntanglement1,EntropicEntanglement2} of quadrature measurements. Recent advances include entropic entanglement criteria based on quasiprobability distributions \cite{EntropicPhaseSpace1,EntropicPhaseSpace2} and measurement-device-independent criteria \cite{PaoloMDIEW}, both of which require performing some form of partial tomography.

The criteria to be chosen would of course be informed by the type of CV system used. The obvious test bed for CV entanglement are optical modes \cite{OpticsCV} and vibrational modes of trapped ions \cite{TrappedIonCV}, but there has also been significant progress for massive oscillators \cite{SarmaOptomechReview}. For many of these CV systems, harmonic dynamics arises naturally in their implementation.

Because of these motivations, we present a criterion to certify the non-Gaussian entanglement of two CV degrees of freedom that exploits the knowledge of the \textit{dynamics} of the systems. We call it a dynamic-based entanglement witness (DEW) \footnote{The name ``dynamical entanglement witness'' would have been more elegant, but could have been confused with a ``witness of \textit{dynamical entanglement}'', which is a notion of entanglement for channels \cite{dynent}. Here we are concerned with entanglement of states}\nocite{dynent}. Specifically, we consider degrees of freedom that undergo types of \textit{harmonic} dynamics, and build on a nonclassicality test for a single oscillator \cite{tsirelson2006often,zaw2022detecting}.

\textit{A priori}, our criterion exhibits two elegant features: the absence of false positives from classical theory, and the fact that only one observable needs to be measured. These features, to be defined precisely below, can be appreciated by contrast with existing entanglement witnesses, based on generalized uncertainty relations \cite{duan2000inseparability,ZhangEtAl,hillery2006entanglement,SU2entanglement, UncertaintyIsPPT1,UncertaintyIsPPT2}. Consider specifically the criterion by \citet{duan2000inseparability}. Generalizing the observables defined in the Einstein-Podolsky-Rosen (EPR) argument \cite{EPR1935}, these authors defined the commuting dimensionless variables $u = \abs{c}\tilde{x}_1 + (1/c)\tilde{x}_2$ (with $\tilde{x}_j=x_j\sqrt{{m_j\omega_j}/{\hbar}}$) and $v = {\abs{c}}\tilde{p}_1 - (1/c)\tilde{p}_2$ (with $\tilde{p}_j=p_j/\sqrt{m_j\hbar\omega_j}$) for some $c \in \mathbb{R}$. The two subsystems are then entangled if $\langle{\pqty{\Delta u}^2}\rangle + \langle{\pqty{\Delta v}^2}\rangle < c^2 + (1/c^2)$. This requires measuring both positions and momenta, and with a precision set by $\hbar$. At the precision of (say) human perception, two springs at equilibrium are described by $\tilde{x}_j=\tilde{p}_j=0$, values which would imply entanglement if plugged naively in the criterion above. Gross though it is, this example shows the danger of false positives.

\textit{A posteriori}, we find that our criterion detects states with negative Wigner functions (thus, non-Gaussian), some of which are missed by all existing criteria. As mentioned earlier, Gaussian states have limited usefulness in many quantum protocols \cite{Positive-Wigner-Simulatable,CV-Error-Correction,CV-Entanglement-Distillation-1,CV-Entanglement-Distillation-2}. In this context, our DEW detects resource states for these protocols \cite{CV-Resource-Theory}. Thus, besides being elegant, our DEW is also a useful addition to the existing toolbox.

\section{The single-oscillator protocol}\label{1D Tsi}

We review the Tsirelson nonclassicality test \cite{tsirelson2006often} following the generalization given in \cite{zaw2022detecting}. The assumption is that the physical quantity $A_1$ is undergoing a uniform precession at pulsation $\omega$, i.e.~$A_1(t)=A_1(0)\cos\omega t+A_2(0)\sin\omega t$, where $A_2$ is another physical quantity. For classical systems, $A_1(t)$ is the value of $A_1$ at time $t$; for quantum systems, it is the corresponding observable in the Heisenberg representation. The protocol for the test (which we call the \textit{precession protocol} hereafter) goes as follows. In each round, the sign of $A_1$ is measured at one of $K$ different times given by $t_k = (k/K)T$, where $K>0$, $k = 0,1,...,K-1$, and $T$ is the period of oscillation. After several rounds, one estimates
\ba\label{eq:defPN}
{P}_K&=&\frac{1}{K}\sum_{k=0}^{K-1}\Bqty{\Pr[A_1(t_k)>0]+\frac{1}{2}\Pr[A_1(t_k)=0]},\,\,\,\,
\ea where the second term in the bracket was introduced in \cite{zaw2022detecting} to avoid singular behaviors for states with noninfinitesimal concentration on $A_1=0$. By inspection \cite{zaw2022detecting}, the upper bound $P_K\leq \mathbf{P}_K^c$ for a classical theory is easily derived: $\mathbf{P}_K^c=1/2$ for $K$ even, and
\ba\label{eq:class Tsi}
\begin{aligned}
    \mathbf{P}_K^c &= \frac{1}{2}\left(1+ \frac{1}{K}\right) \text{ for $K$ odd.}
\end{aligned}
\ea Remarkably, in spite of the fact that the precessing dynamics is identical to the classical one, there exist quantum states for which $P_K > \mathbf{P}_K^c$ for any odd $K>1$.

For the remainder of the Letter, we focus on the \textit{harmonic oscillator}, i.e.~a material point, whose time evolution is governed by the Hamiltonian $H = \frac{1}{2m}p^2 + \frac{1}{2}m\omega^2x^2$. The pair $(A_1,A_2)=(\tilde{x},\tilde{p})$ clearly precesses at pulsation $\omega$ and thus satisfies the assumption. On a given state $\rho$, quantum theory predicts $P_K=\Tr(\rho Q_K)$ where \ba \label{Qop}
Q_K&=&\frac{1}{K}\sum_{k=0}^{K-1} \pos[X(t_k)],
\ea
with $\pos(X)$ defined by $\pos(X)\ket{x} = \frac{1}{2}[1+\sgn(x)]\ket{x}$.

The maximum quantum score $\mathbf{P}_K=\max_{\ket{\psi}}\ev{Q_{K}}{\psi}$ (denoted $\mathbf{P}^\infty_K$ in \cite{zaw2022detecting}) is achieved by $\ket{\mathbf{P}_K}$, the eigenstate of $Q_K$ with the largest eigenvalue. Tsirelson proved that $\mathbf{P}_3\gtrsim 0.709>\mathbf{P}_3^c = 2/3$ \cite{tsirelson2006often}; similar violations are found for all odd $K$ \cite{zaw2022detecting}. The violation can be attributed to having suitable patterns in the Wigner function, in particular, suitably distributed negativities. A state with positive Wigner function cannot give any violation.

\section{Entanglement of two harmonic oscillators}\label{inf dim}

Now, the key insight is that $x$ may be the position of an \textit{effective} oscillator, built out of two (or more) physical ones. We focus on the case of two oscillators, with arbitrary masses and frequencies, and an $x$--$x$ coupling. The standard decomposition in normal modes yields
\ba
    H &=& \sum_{j=1}^2\pqty{\frac{1}{2m_j}p_j^2 + \frac{1}{2}m_j\omega_j^2x_j^2} -
    \frac{1}{2}gx_1x_2 \nonumber\\
    &=&\sum_{\sigma\in \{+,-\}}\frac{1}{2\mu}p_\sigma^2 + \frac{1}{2}\mu\omega_\sigma^2x_\sigma^2, \label{eq:coupled-hamiltonian}
\ea
where $\mu=\sqrt{m_1m_2}$,
\begin{equation}\label{eq:normal-mode-position}
\begin{aligned}
    \!\!\!\!x_+(t) &= \pqty{\frac{m_1}{m_2}}^{1/4}\cos\theta \;x_1(t) + \pqty{\frac{m_2}{m_1}}^{1/4}\sin\theta \;x_2(t) \\
    \!\!\!\!x_-(t) &= \pqty{\frac{m_2}{m_1}}^{1/4}\cos\theta \;x_2(t) -\pqty{\frac{m_1}{m_2}}^{1/4}\sin\theta \;x_1(t),
\end{aligned}
\end{equation}
with mixing angle $\theta = \operatorname{arctan2}\bqty{g,\mu(\omega_1^2-\omega_2^2)}/2$, and normal frequencies $\omega^2_\pm = \frac{\omega_1^2+\omega_2^2}{2} \pm \sqrt{\pqty{\frac{\omega_1^2-\omega_2^2}{2}}^2 + \frac{g^2}{4\mu^2}}$. 
The time evolution of the $x_\sigma(t)$ is a uniform precession around phase space with the period $T_\sigma = 2\pi/\omega_\sigma$. Therefore, the single-oscillator protocol can be performed for coupled oscillators with different frequencies by measuring the normal modes $x_\sigma(t_k+t_0)$ at times $t_k = (k/K) T_\sigma$ for $k=0,1,\dots,K-1$.
There are many ways to estimate $P_K$ for $x_\sigma$: in each round, $x_\sigma$ can be formed from $x_1$ and $x_2$ measured separately, addressed directly (e.g., with motional modes of trapped ions), or measured with an interferometer (e.g., with spatial or polarization modes of photons). 
Notice also that, up to a multiplicative constant, $x_\sigma(t)$ has the same form as $u$ defined in the entanglement criterion by Duan and co-workers \cite{duan2000inseparability}.

It is important to stress that $H$ describes the dynamics during the certification protocol, not the interaction that prepared the state under study. Thus, all values of the parameters are allowed. In particular, the certification protocol can be performed when $g=0$ and $\omega_1=\omega_2$. In this case, $\theta$ can take on any value: indeed, for uncoupled oscillators precessing at the same frequency, all linear combinations of $x_1$ and $x_2$ are normal modes at that same frequency.

Another point to note is that the dynamics are assumed to be known, in which case only one quadrature $x_\sigma$ needs to be measured for the protocol. One could of course replace this assumption by taking the quadratures $\{x_\sigma(t_k)\}_{k=0}^{K-1}$ at the different times to be $K$ different settings of the measurement apparatus. 

If $P_K>\mathbf{P}_K^c$ for $x_\sigma$, the state of that mode has certainly a negative Wigner function. We want to study when one can further infer that the physical subsystems are entangled. This is not straightforward because, by performing the protocol on one of the normal modes, we learn nothing about the state of the other mode: the latter could be very mixed; or the two normal modes may be even entangled. We are going to provide the conditions under which entanglement can indeed be certified.

\section{Results}\label{results}
For the quantum system, we denote the annihilation operators of the two physical oscillators as $\{a_1,a_2\}$: they are the subsystems whose entanglement we want to certify. As hinted, $x_\sigma(t)$ is the position of an effective oscillator denoted by the annihilation operator $a_\sigma$. Specifically, let $\{a_+,a_-\}$ be a new basis of modes, related to the original by the passive transformation
\begin{equation}\label{eq:basis-transformation}
    \pmqty{a_+\\a_-}
    = \pmqty{
        \cos\theta & \sin\theta \\
        -\sin\theta & \cos\theta
    }
    \pmqty{a_1\\a_2}
\end{equation}
with $\theta\in[0,\pi/4]$ the mixing angle previously defined. We are going to study the operator $Q_K$ given by Eq.~\eqref{Qop} for the position operator $X_\sigma(t) = \sqrt{({\hbar}/{2\mu\omega_\sigma})}\pqty{a_\sigma e^{-i\omega_\sigma t} + a_\sigma^\dag e^{i\omega_\sigma t}}$, where $Q_K$ and $X_\sigma$ rely implicitly on $\theta$ via Eq.~\eqref{eq:basis-transformation}.

When $\theta=\pi/4$, we have an analytical proof of existence of entanglement for any violation $P_K > \mathbf{P}_K^{(c)}$:
\begin{result}
If the precession protocol is performed with $\theta=\pi/4$, all states that violate the classical bound \eqref{eq:class Tsi} for $a_+$ (or $a_-$) are entangled in $\{a_1,a_2\}$.
\end{result}
\textit{Proof.} We show this for $a_+$ with a proof by contradiction. The proof for $a_-$ proceeds in a similar way.

Take $\rho=\sum_k p_k \rho_{1}^{(k)} \otimes \rho_{2}^{(k)}$ separable in the $\{a_1,a_2\}$ subsystems. Its Wigner function is
\begin{equation}\label{eq:W(rho_sep_a1a2}
W_{\rho}(\alpha_1,\alpha_2) = \sum_k p_k W_{\rho_{1}}^{(k)}(\alpha_1) W_{\rho_{2}}^{(k)}(\alpha_2),
\end{equation}
where $\{\alpha_1,\alpha_2\}$ are the phase-space coordinates in the $\{a_1,a_2\}$ modes. For $\theta=\pi/4$ in Eq.~\eqref{eq:basis-transformation}, the Wigner function in terms of $\{\alpha_+,\alpha_-\}$, the phase-space coordinates in the $\{a_+,a_-\}$ modes, can be found with a straightforward coordinate transformation:
\begin{equation}
    W_{\rho}(\alpha_+,\alpha_-) = \sum_k p_k W_{\rho_{1}}^{(k)}\!\pqty{
        \frac{\alpha_+ - \alpha_-}{\sqrt{2}}
    } W_{\rho_{2}}^{(k)}\!\pqty{
        \frac{\alpha_+ + \alpha_-}{\sqrt{2}}
    }.
\end{equation}
The measurement outcome in the $a_+$ basis is determined solely by the reduced Wigner function $W_{\tr_-(\rho)}(\alpha_+)=\int\dd[2]{\alpha_-}W_{\rho}(\alpha_+,\alpha_-)$. By converting the passive coordinates in the arguments into active transformations of the states, we find
\ba
W_{\tr_{-}(\rho)}(\alpha_+) &=& 2\sum_k p_k \int\dd[2]{\gamma}W_{\tilde{\rho}_{1}(\alpha_+)}^{(k)}(\gamma)W_{\tilde{\rho}_{2}(\alpha_+)}^{(k)}(\gamma)\nonumber\\\label{eq:convolution}
&=& \frac{2}{\pi}\sum_k p_k \tr(\tilde{\rho}_{1}^{(k)}(\alpha_+)\tilde{\rho}_{2}^{(k)}(\alpha_+)),
\ea
where
\begin{align*}
\tilde{\rho}_{1}^{(k)}(\alpha_+) = &D\!\pqty{\frac{1}{\sqrt{2}}\alpha_+}e^{-i\pi a^\dag a}\rho_{1}^{(k)}e^{i\pi a^\dag a}D^\dag\!\pqty{\frac{1}{\sqrt{2}}\alpha_+}\\
\tilde{\rho}_{2}^{(k)}(\alpha_+) = &D\!\pqty{-\frac{1}{\sqrt{2}}\alpha_+}\rho_{2}^{{(k)}}D^\dag\!\pqty{-\frac{1}{\sqrt{2}}\alpha_+}.
\end{align*}
and $D(\alpha)$ is the usual displacement operator. Thus $W_{\tr_{-}(\rho)} \geq 0$, since it is a convex sum of inner products between density operators \footnote{It was recently pointed out that Eq.~\eqref{eq:convolution} is also known as the \textit{convolution} of two Wigner functions \cite{PositiveWigner,QuantumCentralLimitTheorem}, which has been used in the past to ``smooth out'' the negativities of a Wigner function to define a nonnegative quasiprobability distribution \cite{Smoothed-Wigner-1,Smoothed-Wigner-2}. To our knowledge, this has not been previously exploited to witness entanglement.}\nocite{PositiveWigner,QuantumCentralLimitTheorem,Smoothed-Wigner-1,Smoothed-Wigner-2}. However, negativity in the Wigner function of $a_+$ is necessary for a violation of the classical bound of the precession protocol \cite{zaw2022detecting}. Therefore, when $\theta=\pi/4$, \emph{any} violation of the classical bound witnesses entanglement of the $\{a_1,a_2\}$ subsystems. \hfill \qedsymbol{}

\begin{figure*}[ht!]
    \centering\includegraphics[width=\textwidth]{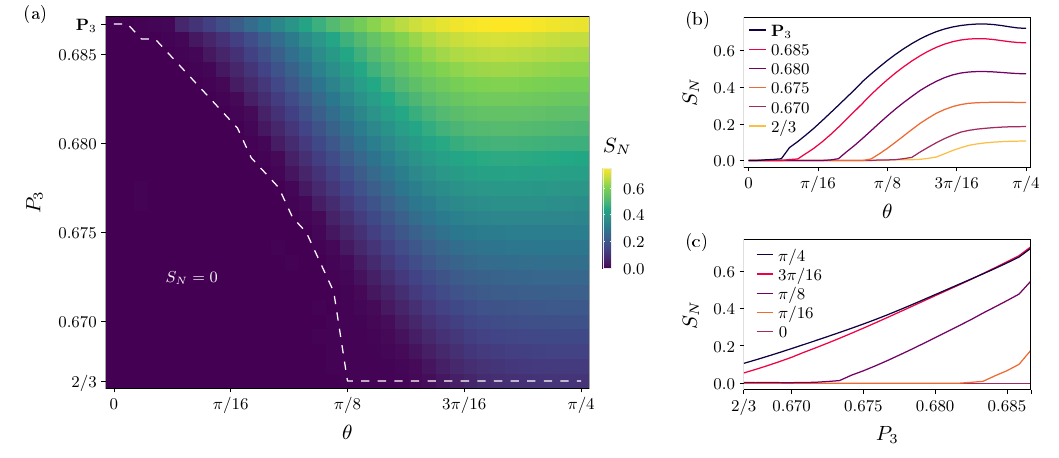}
    \caption{\label{fig:Heatmap of LogNeg}(a) Heat map of logarithmic negativity as a function of the mixing angle $\theta$ and the score $P_3$. Only the range $0 \leq \theta \leq \pi/4$ is shown: other values of $\theta$ corresponds to this range with a sign change of $x_1$ and $x_2$, which can be effected with a local unitary on the $\{a_1,a_2\}$ basis and hence does not affect the amount of entanglement. The dimension of the Hilbert space is truncated at $\mathbf{n}=11$ for both modes $a_+$ and $a_-$, and $\mathbf{P}_3$ is the maximum quantum score under this truncation. The dotted line separates the states with $S_N>0$ (such that $S_N-\varepsilon > 0$, where $\varepsilon {}\lesssim 10^{-4}$ is the dual gap of the SDP) and those with $S_N=0$. (b) Horizontal line cuts of the heat map. For fixed $P_3$, $S_N$ increases with $\theta$. (c) Vertical line cuts of the heat map. For fixed $\theta$, $S_N$ increases with $P_3$.}
\end{figure*}

Next, we are going to study in detail the protocol with $K=3$. For any value of $\theta$ and of $P_3 > \mathbf{P}_3^c=2/3$, we are going to compute a numerical lower bound on the amount of certifiable entanglement of $\{a_1,a_2\}$. We choose the logarithmic negativity $S_N(\rho) = \log\tr\!\abs{\rho^{\Gamma_{2}}}$ as the quantifier of entanglement. Since
$\min_{\rho} S_N(\rho)=\log \min_{\rho}\tr\!\abs{\rho^{\Gamma_{2}}}$, and
\begin{equation}
\begin{aligned}
    &\min_{\rho} \tr\!\abs{\rho^{\Gamma_{2}}} \\
    &\;\;\text{subject to}\;\tr[\rho Q_3(\theta)] = P_3
\end{aligned}
\end{equation}
is a minimization of the trace norm under convex constraints, it can be cast as a standard semidefinite program (SDP) when $\rho$ is truncated in the basis of the excitations of $a_+$ and $a_-$ \cite{SDP-nuclear-norm}.

We run the SDP for truncation $0 \leq n_{+},n_{-} \leq \mathbf{n}=11$. Both the form of the SDP and the choice of the truncation are described in \cite{SM_DEW}\nocite{convexjl,convex-textbook,Gell-Mann-Matrices,almost-always-entangled,Two-Party-Cat-1,Two-Party-Cat-2,optimal-entanglement-witness}, and the script used to perform the SDP is available at \cite{SDP-script}. The results are plotted in Fig.~\ref{fig:Heatmap of LogNeg}. We do not fully understand the dependence of the logarithmic negativity on $\theta$ and $P_3$ due to the complexity of the states involved. Broadly speaking, what we do observe from Fig.~\ref{fig:Heatmap of LogNeg}(b) is that the certifiable $S_N$ increases with $\theta$ and $P_3$, as shown more explicitly by the line cuts. For fixed values of $P_3 \sim \mathbf{P}_3$, where $\mathbf{P}_3$ is the maximum quantum score under the truncation $\mathbf{n}$, the certifiable $S_N$ increases with $\theta$ until a peak around $\theta \gtrsim 3\pi/16$. Afterwards, the entanglement decreases for larger values of $\theta$, although only slightly. In practice, $\theta$ is determined by the system, and one would refer to the corresponding line cut in Fig.~\ref{fig:Heatmap of LogNeg}(c). There, for fixed values of $\theta$, we find that the entanglement monotonically increases with $P_3$. 

We already knew that every $P_3$ certifies entanglement when $\theta={\pi}/{4}$, and the graph indicates that this remains true down to $\theta\simeq {\pi}/{8}$. Below this value, one needs a sufficiently large $P_3$, a low violation of the classical bound being compatible with separable states. When $\theta=0$, the precession protocol is performed on the first oscillator, and so no amount of violation detects entanglement.

\section{Comparison with other witnesses}\label{discuss}

Now we put our DEW in the context of entanglement witnesses for continuous variables (CVs), by comparing it to other criteria.

First of all, our DEW uses \emph{quadrature measurements} in the terminology of quantum optics. Other measurements than quadratures can be used to witness CV entanglement: for instance, one witness in \citet{ZhangEtAl} uses local measurements of the generators of $\mathrm{SU}(N)$. In fact, any CV entanglement can, in principle, be witnessed by projecting the state into a finite dimensional subspace, then applying techniques to witness entanglement of qudits \cite{FiniteEnoughForCV}. Quadrature measurements have the appeal of having a classical analog, are practical in many platforms, and in some setups may be even the only available ones at this time (e.g.,~optomechanical systems with large masses). 

As already mentioned in the introduction, the other entanglement witnesses we are aware of are open to \textit{false positives} from classical theory \cite{duan2000inseparability,ZhangEtAl,hillery2006entanglement, UncertaintyIsPPT1,UncertaintyIsPPT2}. By contrast, in the case of our DEW, poor precision or wrong calibration may prevent the detection of entanglement, but will not lead to false positives. This is similar to what happens with Bell inequalities, where noise and lack of precision may decrease or cancel the violation, but not fake it.

Having mentioned this, it is natural to move on to the comparison in terms of characterization of the devices. Fully device-independent entanglement witnesses (Bell inequalities) that use only quadrature measurements have been hard to find: the few known examples are very specific \cite{BWBell,NhaBell,GrangierBell}. Recently, a measurement-device-independent criterion was introduced, under the assumption that a trusted source of coherent states is available \cite{PaoloMDIEW}. Meanwhile, our DEW is \textit{semi-device-independent}: it works under the assumptions that the dynamics is a uniform precession, and that the same quadrature is measured whatever time is picked. Both assumptions are well defined both in classical and in quantum theory. In fact, the notions of ``position'' and ``uniform precession'' are operational, and their meaning immediate by everyday experience (what may not be immediate is the identification of a normal mode).

Lastly, let us compare the states that are detected by the existing criteria against those detected by our DEW. Many existing criteria can detect entangled Gaussian states, like the two-mode squeezed state and its limiting case, the EPR state; our DEW misses these states, since their Wigner function is positive. Our DEW is also unsuitable for states with even rotational symmetry, like photon-subtracted or added squeezed number states \cite{Photon-Subtraction-1,Photon-Subtraction-2} and spin coherent states \cite{SU2entanglement}, because they commute with the total parity operator (details in \cite{SM_DEW}). Conversely, for any $K\geq 3$ odd, consider the states
\ba\label{eq:OddRotDetected}
\ket{\{\psi_n\}_{\mathbf{n}}}&=& \sum_{j=0}^{\mathbf{n}K} \ket{\Psi^j}_1\otimes\ket{j}_{2}\\&=& \Big(\sum_{n=0}^{\mathbf{n}}\psi_n\ket{n{K}}_{+}\Big)\otimes\ket{0}_{-}\equiv \ket{\Psi_\mathbf{n}}_{+}\otimes\ket{0}_{-}\nonumber
\ea with
\ban
\ket{\Psi^j}&=&\sum_{n=\lceil {j}/{K} \rceil}^{\mathbf{n}} \psi_n
        \sqrt{\pmqty{nK\\j}}\pqty{\cos\theta}^{nK-j}\pqty{\sin\theta}^j \ket{nK-j}
\ean
where $\mathbf{n}$ can go to infinity, and the pairs of modes are related by Eq.~\eqref{eq:basis-transformation}. These states exhibit odd rotational symmetry---making them candidates for a type of bosonic error correcting codes \cite{Bosonic-Code}---and are entangled as long as $\theta\bmod\pi/2\neq 0$ and $\abs{\psi_{{n}}} \neq \delta_{n,n_0}$ for one value $n_0$. All are missed by \cite{duan2000inseparability,ZhangEtAl,PaoloMDIEW}, and some also by \cite{hillery2006entanglement, SU2entanglement} (see \cite{SM_DEW} for details). Clearly our DEW detects the entanglement of all the states \eqref{eq:OddRotDetected} such that $\ket{\Psi_{\textbf{n}}}$ violates the original precession protocol with $K$ possible probing times. In particular, the eigenstates of $Q_{K}$ with maximal eigenvalue are of the form $\ket{\Psi_{\textbf{n}}}$, and so the corresponding $\ket{\{\psi_n\}_{\mathbf{n}}}$ is optimally detected by our DEW. As another example: for $K=3$, $\theta=\pi/4$ and a suitable choice of the $\psi_n$ (see \cite{SM_DEW}), the state \eqref{eq:OddRotDetected} is the entangled three-level cat state
\ban
\ket{\Psi(\alpha)}\propto \sum_{k=-1}^{1} \ket{\alpha e^{i2\pi k/3}}_1 \otimes \ket{\alpha e^{i2\pi k/3}}_2.
\ean This state is detected by our DEW for $0.88\lesssim\abs{\alpha}\lesssim 1.23$, while for $1.23\lesssim\abs{\alpha}\lesssim 1.82$ it is detected by the criteria of \cite{hillery2006entanglement, SU2entanglement}. Finally, let us notice that one does not need very high excitations: our DEW with $K=3$ detects the state with $(\psi_0,\psi_1,\psi_2)\simeq (0.6172, -0.7017, 0.3450)$, which is a superposition of 0, 3, and 6 excitations in $a_+$.  

More generally, our DEW is not a subset of any member of the family of uncertainty-based entanglement witnesses defined by \citet{UncertaintyIsPPT1} and \citet{UncertaintyIsPPT2}, which include \cite{duan2000inseparability, hillery2006entanglement,SU2entanglement} as special cases.

\section{Conclusion}\label{conc}

We have introduced a dynamic-based entanglement witness for two harmonic oscillators. It consists of certifying the quantumness of a normal mode using the Tsirelon protocol \cite{tsirelson2006often,zaw2022detecting}: the entanglement of the physical oscillators can then be inferred, without having any information about the other normal mode (obviously, having also some information about it can only tighten the lower bounds that we have obtained).

Our criterion detects a different set of states than those captured by previous ones. Also, it does not rely on other features of quantum theory (e.g.,~some form of uncertainty relations): it only assumes the form of the dynamics. As only straightforward coordinate measurements are used, and false positives from classical theory are excluded, our criterion is useful for objects which are too massive to be fully tomographed. It is timely, as recent advances in optomechanics have allowed for quantum control of objects with masses in the mesoscopic and macroscopic scales \cite{SarmaOptomechReview}. The ability to generate and detect quantum effects in such systems has proven useful in technologies that aim to exploit the effects of quantum mechanics, in particular, quantum sensing \cite{QMFreeSubsys, StationaryOptoVienna} and metrology \cite{CVMetrology}. Moreover, it is also important in experimental studies of fundamental physics, in tests of collapse theories \cite{entanglement-test-of-collapse-theories} and quantum-classical transitions \cite{quantum-classical-transition}. 

In order to focus on the essentials of the idea, in this Letter we have kept to the simplest form of dynamics, that of two coupled harmonic oscillators. The underlying Tsirelson protocol can be extended to a wide class of Hamiltonians, possibly with an additional energy constraint \cite{anharmonic}. Dynamic-based entanglement certification can also be extended to dissipative dynamics: leaving quantitative estimates for future work, it is clear that one will still be able to certify entanglement provided the relaxation time is long enough. We have also only focused on the bipartite scenario here, but an obvious future direction would be to extend our protocol for the multipartite case. As an early example, we show in  \cite{SM_DEW} that by performing the precession protocol in the $\propto a_1+a_2+a_3$ mode, our DEW can detect genuine tripartite entanglement \cite{Genuine-Multipartite-Entanglement}.

\section*{Acknowledgments}

We acknowledge financial support from the National Research Foundation and the Ministry of Education, Singapore, under the Research Centres of Excellence programme. We also thank the National University of Singapore Information Technology for the use of their high performance computing resources.

\bibliography{refs}

\clearpage

\phantomsection\addcontentsline{toc}{part}{Supplemental Material for: Dynamics-Based Entanglement Witnesses for Non-Gaussian States of Harmonic Oscillators}
\title{Supplemental Material for: Dynamics-Based Entanglement Witnesses for Non-Gaussian States of Harmonic Oscillators}

\maketitle

\setcounter{section}{0}
\setcounter{figure}{0}
\setcounter{equation}{0}
\renewcommand{\thesection}{S\arabic{section}}
\renewcommand{\thefigure}{S\arabic{figure}}
\renewcommand{\theequation}{S\arabic{equation}}

\section{\label{apd:SDP}Semidefinite programming}
In this section, we detail the numerical procedures performed to obtain the results reported in the main text using semidefinite programming (SDP). An SDP is an optimization of the form
\begin{equation}\label{eq:SDP-standard-form}
    \min_{\{x_k\}_k} \sum_k a_k x_k, \;\;\text{subject to}\;\; B_0 + \sum_k x_k B_k \succcurlyeq  0,
\end{equation}
where $\{a_k\}_k$ are real numbers, $\{B_k\}_k$ are Hermitian matrices, and $\{x_k\}_k$ are real variables over which we perform the optimization. Note that linear constraints of the form $\sum_k c_k x_k = c_0$ for real numbers $\{c_k\}_k$ and $\sum_k D_k x_k = D_0$ for Hermitian matrices $\{D_k\}_k$ belong to the type of constraint given in Eq.~\eqref{eq:SDP-standard-form}.

Once a problem has been placed into the standard form given in Eq.~\eqref{eq:SDP-standard-form}, there are many SDP solvers available that can be used to perform the optimization. We used \texttt{Convex.jl} for this paper \cite{convexjl}, and the script used to generate the results is available at \cite{SDP-script}. As SDP is a form of convex optimization that returns the global optimum up to a specified precision \cite{convex-textbook}, the numerical results obtained in this section are rigorous under the truncation justified in Section~\ref{apd:truncation-justification}.

\subsection{Minimizing entanglement for observed \texorpdfstring{$P_K$ }{violation of classical bound} and given \texorpdfstring{$\theta$}{mixing angle}}
The entanglement of a state $\rho$ in the $\{a_1,a_2\}$ basis can be quantified with the logarithmic negativity $S_N(\rho) = \log\tr\!\abs{\rho^{\Gamma_{2}}}$. By bounding the logarithmic negativity from below, we can deduce the minimum entanglement of a state that violates the classical bound of the precession protocol.

To cast the minimization of $S_N(\rho)$ \textcolor{black}{into the standard form of an SDP given in Eq.~\eqref{eq:SDP-standard-form}}, we introduce a Hermitian basis of operators $\{B_j\}_{j=0}^{\pqty{\mathbf{n}+1}^2-1}$ for both $a_\sigma$ systems, for some truncation $0\leq n_{\sigma} \leq \mathbf{n}$, with the convention that $B_0 \propto \mathbb{1}$ and $\tr(B_jB_k) = \delta_{j,k}$. In our numerical program, we used as $B_j$ the normalized form of the generalized Gell-Mann matrices \textcolor{black}{\cite{Gell-Mann-Matrices}}. The operator basis over the truncation of $\{a_+,a_-\}$ is $\{B_j \otimes B_k\}_{j,k=0}^{\pqty{\mathbf{n}+1}^2-1}$. Excluding $B_0 \otimes B_0$, we collect them into a vector as
\begin{equation*}\begin{aligned}
\overrightarrow{B\!B} \equiv \Large(
    &\phantom{B_1\otimes B_0,}B_0 \otimes B_1,\;\dots,\;B_0\otimes B_{\pqty{\mathbf{n}+1}^2-1},\\
    &B_1\otimes B_0,B_1\otimes B_1,\;\dots,\;B_1\otimes B_{\pqty{\mathbf{n}+1}^2-1},\\
    &\phantom{B_1\otimes B_0,B_1\otimes B_0,}\;\dots,\;B_{\pqty{\mathbf{n}+1}^2-1} \otimes B_{\pqty{\mathbf{n}+1}^2-1}\Large).
\end{aligned}\end{equation*}
Then $\rho = {B_0 \otimes B_0}/{\tr(B_0)^2} + \vec{x}\cdot\overrightarrow{B\!B}$
and \ba \rho^{\Gamma_{2}} = \frac{\pqty{B_0 \otimes B_0}^{\Gamma_{2}}}{\tr(B_0)^2} + \vec{x}\cdot\overrightarrow{B\!B}^{\Gamma_{2}},\ea where $\vec{x} = {\tr}(\rho \overrightarrow{B\!B})$, and $\overrightarrow{B\!B}^{\Gamma_{2}}$ is found by representing each operator in $\overrightarrow{B\!B}$ as a matrix in the $\{a_1,a_2\}$ basis (here is where $\theta$ is used) and taking the partial transpose in the usual way. Denoting its positive eigenvalues as $\lambda_{+k}$ and its negative eigenvalues as $-\lambda_{-k}$, so that $\lambda_{\pm k} \geq 0$, we have the spectral decomposition
\begin{equation}
\begin{aligned}
    \rho^{\Gamma_{2}} &= \underbrace{\sum_{\lambda_{+k}}\lambda_{+k}\ketbra{\lambda_{+k}}}_{\equiv \varrho_+}
    - \underbrace{\sum_{\lambda_{-k}}\lambda_{-k}\ketbra{\lambda_{-k}}}_{\equiv \varrho_-} \\
    &= \varrho_+ - \varrho_-,\qquad\text{where}\;\varrho_{\pm} \succcurlyeq 0.
\end{aligned}
\end{equation}
Considering that the restriction $0 \leq n_+,n_-\leq \mathbf{n}$ allows for states in the $\{a_1,a_2\}$ basis with $0 \leq n_{1},n_{2}\leq 2\mathbf{n}$, the Hermitian operator basis for each $a_i$ is $\{A_j\}_{j=0}^{\pqty{2\mathbf{n}+1}^2-1}$. The full space of $\varrho_{\pm}$ is spanned by $\{A_j\otimes A_k\}_{j,k=0}^{\pqty{2\mathbf{n}+1}^2-1}$, and we similarly define $\overrightarrow{A\!A} \equiv (A_0\otimes A_1,\dots,A_{\pqty{2\mathbf{n}+1}^2-1}\otimes A_{\pqty{2\mathbf{n}+1}^2-1})$. Then,
\begin{equation}
\begin{aligned}
    \varrho_+ &= z\frac{A_0\otimes A_0}{\tr(A_0)^2} + \vec{y}_+\cdot\overrightarrow{A\!A} \\
    \varrho_- &= (z-1)\frac{A_0\otimes A_0}{\tr(A_0)^2} + \vec{y}_-\cdot\overrightarrow{A\!A},
\end{aligned}
\end{equation}
where $\vec{y}_{\pm} = \tr(\varrho_{\pm}\overrightarrow{A\!A})$, $z = \tr(\varrho_+)$, and $\tr(\varrho_-)$ is fixed by $\tr(\varrho_+-\varrho_-) = 1$. This means that the logarithmic negativity of $\rho$ in the $\{a_1,a_2\}$ basis is
\begin{equation}
    S_N(\rho) = \log\tr\!\abs{\varrho_+-\varrho_-}
    = \log(2z-1).
\end{equation}
Finally, defining $\vec{q} \equiv \tr(Q_K\overrightarrow{B\!B})$, we have
\begin{equation}
P_K = \tr(\rho Q_K) = \frac{1}{2} + \vec{x}\cdot\vec{q}.
\end{equation}

\subsubsection*{SDP Procedure}
Define $A_j$ and $B_j$ according to the preceding discussion. Then,
\begin{equation}
\begin{aligned}
    &\min_{\begin{array}{rcl}
        \scriptstyle\vec{x} &\scriptstyle\in& \scriptstyle\mathbb{R}^{\pqty{\mathbf{n}+1}^4-1}; \\
        \scriptstyle\vec{y}_+,\vec{y}_- &\scriptstyle\in& \scriptstyle\mathbb{R}^{\pqty{2\mathbf{n}+1}^4-1}; \\
        \scriptstyle z&\scriptstyle\in&\scriptstyle\mathbb{R}
    \end{array}} z \\
    &\;\;\text{subject to}\\
    &\;\;\;\;P_K = \tfrac{1}{2} + \vec{x}\cdot\vec{q} \\
    &\;\;\;\;\tfrac{\pqty{B_0 \otimes B_0}^{\Gamma_{2}}}{\tr(B_0)^2} + \vec{x}\cdot\overrightarrow{B\!B}^{\Gamma_{2}} = \tfrac{A_0\otimes A_0}{\tr(A_0)^2} + \pqty{\vec{y}_+-\vec{y}_-}\cdot\overrightarrow{A\!A}, \\[3ex]
    &\;\;\;\;\spmqty{
         \frac{B_0 \otimes B_0}{\tr(B_0)^2} + \vec{x}\cdot\overrightarrow{B\!B} & 0 & 0 \\
         0 & z\frac{A_0\otimes A_0}{\tr(A_0)^2} + \vec{y}_+\cdot\overrightarrow{A\!A} & 0 \\
         0 & 0 & (z-1)\frac{A_0\otimes A_0}{\tr(A_0)^2} + \vec{y}_-\cdot\overrightarrow{A\!A}
    } \succcurlyeq 0.
\end{aligned}
\end{equation}
With the minimum $z$, $S_N(\rho) = \log(2z-1)$. For the results in the main text, we used $K=3$ and $\mathbf{n}=11$. The choice of $\mathbf{n}$ is justified in the following section.

\subsection{Choice of truncation \texorpdfstring{$\mathbf{n}$}{}\label{apd:truncation-justification}}
\textcolor{black}{The numerical results from the SDP optimization is rigorous under the energy constraint $E_{\pm} \leq \hbar\omega_{\pm}(\mathbf{n}+\frac{1}{2})$, where $E_{\pm}$ is the energy of the system in the normal mode. The choice of the truncation $\mathbf{n}$ would be informed by the energy scale of the experimental setup: for example, if only the first few energy levels of the harmonic oscillator of order $n_\pm \sim 1$ will be accessed, it is adequate to choose a truncation $\mathbf{n} \sim 10$.}

\begin{figure}[hb!]
    \centering
    \includegraphics[width=\columnwidth]{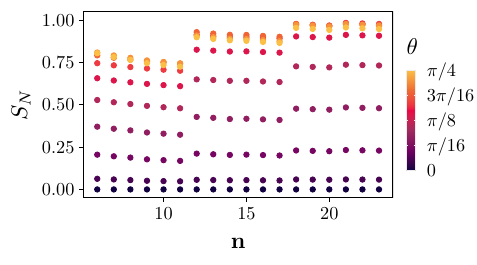}
    \caption{\label{fig:compare all mixed b2}Log negativity of maximally-violating states against truncation $\mathbf{n}$ for various values of $\theta$, found with SDP optimization. In all cases, the minimum $S_N$ occurs at $\mathbf{n}=11$.}
\end{figure}

To choose a representative choice of $\mathbf{n}$ for Fig.~\ref*{fig:Heatmap of LogNeg} in the main text, we consider how our choice of truncation affects the entanglement of states that maximally violate the classical bound. For each $\mathbf{n}$, we perform an SDP optimization as laid out in the preceding section with $P_3 = \mathbf{P}_3$, where $\mathbf{P}_3$ is the largest eigenvalue of $Q_3$ restricted to the truncated subspace $0 \leq n_{+},n_{-} \leq \mathbf{n}$.

In Fig.~\ref{fig:compare all mixed b2}, we plot the logarithmic negativity against the truncation $\mathbf{n}$ for various values of $\theta$. We find that the log negativity increases with $\mathbf{n}$ in steps of $\Delta\mathbf{n} = 6$. This reflects how $\mathbf{P}_K$ increases in steps of $\Delta\mathbf{n} = 2K$, which is due to the symmetry of the protocol and the relationship between the odd and even number states of $\ket{\mathbf{P}_K}$ \cite{zaw2022detecting}. Importantly, the minimum value of $S_N$ is found for $\mathbf{n}=11$ for all values of $\theta$. As such, choosing $\mathbf{n}=11$ for the numerical results in our main text provides us with a conservative estimate of the amount of entanglement witnessed for a given violation $P_3$.

\section{Class of states missed by our DEW}
The total parity operator $\Pi$ satisfies $\Pi^2 = \mathbbm{1}$ and acts on the position operator as $\Pi X_\mu(t) \Pi = - X_\mu(t)$, where $\mu \in \{1,2,+,-\}$. If $\comm{\rho}{\Pi} = 0$, then
\begin{align}
    \ev{\sgn[X_+(t)]} &= \tr{\sgn[X_+(t)]\rho} \nonumber\\
    &= \tr{\sgn[X_+(t)]\Pi\rho\Pi} \nonumber\\\label{eq:parity-states}
    &= \tr{\Pi\sgn[X_+(t)]\Pi\rho}\\
    &=\tr\{\sgn[-X_+(t)]\rho\} \nonumber\\
    &= -\ev{\sgn[X_+(t)]}. \nonumber
\end{align}
Equation~\eqref{eq:parity-states} implies $\ev{\sgn[X_+(t)]} = 0$. Using the definition of $Q_K$ from Eq.~\eqref{Qop}, $P_K = \tr(Q_K\rho) = 1/2$. Therefore, any state that commutes with the total parity operator cannot be detected by our DEW.

In particular, this includes spin coherent states \cite{SU2entanglement} and photon-subtracted squeezed number states \cite{Photon-Subtraction-1, Photon-Subtraction-2} of the form \begin{equation*}\ket{\psi} = a_1^{m_1}a_2^{m_2}S_{2}(\zeta)\pqty\big{\ket{n_1}_1\otimes\ket{n_2}_2},
\end{equation*}
where $S_{2}(\zeta)={\exp}(\zeta^* a_1 a_2 - \zeta a_1^\dag a_2^\dag)$ is the two-mode squeezing operator. As $\Pi\ket{\psi} = (-1)^{m_1+m_2+n_1+n_2}\ket{\psi}$, $\comm{\ketbra{\psi}}{\Pi} = 0$. A similar proof holds for photon-added squeezed number states.

\section{\label{apd:DuanHillMiss}Class of states missed by uncertainty-based entanglement witnesses}

For $K\geq3$ odd, consider the family of states that was denoted $\ket{\{\psi_n\}_{\mathbf{n}}}$ in the main text, here labelled differently for conciseness:
\ba
\ket{\Phi_{\mathbf{n}}}&=& \sum_{j=0}^{\mathbf{n}K} \ket{\Psi^j}_1\otimes\ket{j}_{2} \label{eq:statesmissedapp}\\
    &=& \Big(\sum_{n=0}^{\mathbf{n}}\psi_n\ket{n \textcolor{black}{K}}_{+}\Big)\otimes\ket{0}_{-}\equiv \ket{\Psi_\mathbf{n}}_{+}\otimes\ket{0}_{-},\nonumber
\ea
where the relation between the pairs of modes is given in Eq.~(6) of the main text, $\mathbf{n}$ can go to infinity, and where
\ban
\ket{\Psi^j}&=&\sum_{n=\lceil {j}/{K} \rceil}^{\mathbf{n}} \psi_n
        \sqrt{\spmqty{nK\\j}}\pqty{\cos\theta}^{nK-j}\pqty{\sin\theta}^j \ket{nK-j}.
\ean
As these states are separable in the $\{a_+,a_-\}$ basis, and do not take the form of the states given in \citet{almost-always-entangled}, they are entangled in the $\{a_1,a_2\}$ basis provided $\theta\bmod\pi/2\neq 0$ and $\abs{\psi_{{n}}} \neq \delta_{n,n_0}$ for one value $n_0$.

Notice that these states are a linear combination of number states that are multiples of $K$. Such states are symmetric over a $2\pi/K$-rotation, and belong to a class of bosonic error-correcting codes \cite{Bosonic-Code}. The states in this family that are detected by our DEW are such that the reduced state $\ket{\psi_\mathbf{n}}_{+}$ violates the original precession protocol. For $K=3$, this includes the maximally-violating states of $Q_3$ under the truncation $3\mathbf{n}$ and $\psi_n \propto {(\sqrt{2}\alpha)^{3n}}/{\sqrt{\pqty{3n}!}}$ for certain values of $\alpha$. Explicit examples of these states can be found in Section~\ref{apd:SD-criteria}.

For these states, we have
\begin{subequations}\label{eq:expectations}
\begin{align}\label{eq:ev_a1}
    \ev{a_1}{\Phi_{\mathbf{n}}} &=
    \ev{a_1^2}{\Phi_{\mathbf{n}}} = 0,\\\label{eq:ev_a2}
    \ev{a_2}{\Phi_{\mathbf{n}}} &=
    \ev{a_2^2}{\Phi_{\mathbf{n}}} =
    0, \\\label{eq:ev_a1a2}
    \ev{a_1 a_2}{\Phi_{\mathbf{n}}} &= 0, \\\label{eq:ev_a1a1}
    \ev{a_1^\dag a_1}{\Phi_{\mathbf{n}}} &= \pqty{\cos\theta}^2\underbrace{\ev{a_+^\dag a_+}{\Psi_\mathbf{n}}}_{\equiv \ev{n}_{\Psi_{\mathbf{n}}}},\\\label{eq:ev_a2a2}
    \ev{a_2^\dag a_2}{\Phi_{\mathbf{n}}} &= \pqty{\sin\theta}^2\ev{n}_{\Psi_{\mathbf{n}}}, \\\label{eq:ev_a1a2t}
    \ev{a_1^\dag a_2}{\Phi_{\mathbf{n}}} &=
    \sin\theta\cos\theta \ev{n}_{\Psi_{\mathbf{n}}}\\\label{eq:ev_a1a1a2a2}
    \ev{a_1^{\dag 2} a_2^2}{\Phi_{\mathbf{n}}}
    &=\ev{a_1^\dag a_1 a_2^\dag a_2}{\Phi_{\mathbf{n}}} \\
    &= \pqty{\sin\theta\cos\theta}^2\pqty{\ev{n^2}_{\Psi_\mathbf{n}} - \ev{n}_{\Psi_\mathbf{n}}}.\nonumber
\end{align}
\end{subequations}

\subsection{Application to criterion by Duan \emph{et al.}}
In the criterion laid out by \citet{duan2000inseparability}, a state is entangled if for some $c\in \mathbb{R}$,
\begin{equation}\label{eq:Duan Crit}
\ev{\pqty{\Delta u}^2} + \ev{\pqty{\Delta v}^2} < c^2 + \frac{1}{c^2},
\end{equation}
where
\begin{align*}
    u &= \abs{c}\sqrt{\frac{m_1\omega_1}{\hbar}}x_1 + \frac{1}{c}\sqrt{\frac{m_2\omega_2}{\hbar}}x_2 \\
    &= \frac{\abs{c}}{\sqrt{2}}\pqty{a_1 + a_1^\dag} + \frac{1}{c\sqrt{2}}\pqty{a_2+a_2^\dag},\\
    v &= \frac{\abs{c}}{\sqrt{m_1\omega_1\hbar}}p_1 - \frac{1}{c\sqrt{m_1\omega_1\hbar}}p_2\\
    &= \frac{\abs{c}}{\sqrt{2}i}\pqty{a_1 - a_1^\dag} + \frac{1}{c\sqrt{2}i}\pqty{a_2 - a_2^\dag}.
\end{align*}
For the class of states in \textcolor{black}{Eq.~}\eqref{eq:statesmissedapp}, \textcolor{black}{with Eq.~\eqref{eq:expectations},} we have
\begin{align*}
    &\ev{\pqty{\Delta u}^2} + \ev{\pqty{\Delta v}^2} \\
    &= \sin(2\theta)\ev{n}_{\psi_\mathbf{n}}\pqty{
        \frac{c^2}{\tan\theta} + \frac{\tan\theta}{c^2} +
        \sgn(c)
    } + c^2+\frac{1}{c^2}.
\end{align*}
Note that $\sin\theta,\cos\theta > 0$ for the valid range of $\theta$. Using $\sgn(x) \geq -1$ and $x+1/x \geq 2$ for $x\geq0$,
\begin{equation}
\begin{aligned}
    \ev{\pqty{\Delta u}^2} + \ev{\pqty{\Delta v}^2} &\geq \sin(2\theta)\ev{n}_{\Psi_\mathbf{n}} + c^2+\frac{1}{c^2} \\
    &> c^2+\frac{1}{c^2},
\end{aligned}
\end{equation}
since $\abs{\psi_0}^2 \neq 1 \implies \ev{n}_{\Psi_\mathbf{n}} > 0$.

Therefore, all states in this family fails the Duan criterion for any choice of $c$, while some---in particular, the maximally-violating state with or without truncation---are detected by our criterion.

\subsection{Application to criterion by Zhang et al.}
For states with $\ev{x_j}=\ev{p_j} = 0$, as is the case for the family of states given in \eqref{eq:statesmissedapp}, the entanglement criteria given in (6) of \citet{ZhangEtAl} becomes
\begin{equation}\label{eq:zhang-simplified}
\begin{aligned}
    4\ev{a_1^\dag a_1}\ev{a_2^\dag a_2} &< 4\abs{\ev{a_1a_2}}^2, \\
    4\ev{a_1^\dag a_1}\ev{a_2^\dag a_2} &< 4\abs{\ev{a_1^\dag a_2}}^2,
\end{aligned}
\end{equation}
where the state is entangled if any of the two inequalities are met. For our family of states, \textcolor{black}{with Eq.~\eqref{eq:expectations},} we have
\begin{equation}
\begin{gathered}
    4\ev{a_1^\dag a_1}\ev{a_2^\dag a_2} = 4\abs{\ev{a_1^\dag a_2}}^2
    = \pqty{\sin(2\theta)}^2\ev{n}^2_{\Psi_{\mathbf{n}}},\\
    4\abs{\ev{a_1a_2}}^2 = 0.
\end{gathered}
\end{equation}
Hence, neither of the inequalities in Eq.~\eqref{eq:zhang-simplified} are satisified, so no member of our family of states will be found to be entangled by {\protect\NoHyper\citeauthor{ZhangEtAl}\protect\endNoHyper}'s criteria.

\subsection{Application to criterion by Abiuso et al.}
\citet{PaoloMDIEW} introduces a trusted source of coherent states $\ket{\alpha}_{\alpha}$ and $\ket{\beta}_{\beta}$, in the ancilla modes $a_\alpha$ and $a_\beta$ respectively. This source produces coherent states following the distribution
\begin{equation}
    P(\alpha) = \frac{1}{\pi \sigma^2}e^{-{\abs{\alpha}^2}/{\sigma^2}},\quad P(\beta) = \frac{1}{\pi \sigma^2}e^{-{\abs{\beta}^2}/{\sigma^2}}.
\end{equation}
Then, a pure state of the total system is of the form $\ket{\Psi_{\text{all}}} = \ket{\alpha}_{\alpha}\otimes\ket{\psi}_{1,2}\otimes\ket{\beta}_{\beta}$, where the entanglement of $\ket{\psi}_{1,2}$ over the $\{a_1,a_2\}$ basis is of interest.

This criterion can detect in principle any entangled state, if no restriction is put on the measurements. When restricted to quadratures, the criterion becomes similar to the one of \citet{duan2000inseparability} and can detect Gaussian entangled states. Let us now prove that it misses the family of states under consideration here. Given the pair of observables

\ban
    U_\kappa &=& \frac{\kappa}{2}\pqty{a_\alpha + a_\alpha^\dag + a_1 + a_1^\dag} - \frac{1}{2\kappa}\pqty{a_\beta + a_\beta^\dag + a_2 + a_2^\dag} \\
    && {}-{}\frac{1}{\sqrt{2}}\pqty{\kappa\Re(\alpha) - \frac{\Im(\beta)}{\kappa}},\\
    V_\kappa &=& \frac{\kappa}{2i}\pqty{a_\alpha - a_\alpha^\dag - a_1 + a_1^\dag} + \frac{1}{2i\kappa}\pqty{a_\beta - a_\beta^\dag - a_2 + a_2^\dag} \\
    && {}-{}\frac{1}{\sqrt{2}}\pqty{\kappa\Re(\alpha) + \frac{\Im(\beta)}{\kappa}},
\ean
the criterion by \citet{PaoloMDIEW} states that $\ket{\psi}_{1,2}$ is entangled if
\begin{equation}
\ev{U_\kappa^2} + \ev{V_\kappa^2} < \frac{\kappa^2 + \kappa^{-2}}{2}\frac{\sigma^2}{1+\sigma^2}.
\end{equation}
Here, the expectation value of $U_\kappa^2$ is over the distribution of the coherent states
\begin{equation}
\ev{U_\kappa^2} = \int\dd[2]{\alpha}\int\dd[2]{\beta} P(\alpha)P(\beta)\ev{U_\kappa^2}{\Psi_{\text{all}}},
\end{equation}
with a similar expression for $V_\kappa^2$.
\begin{widetext}
Replacing $\ket{\psi}_{1,2}$ with the states $\ket{\Psi_{\mathbf{n}}}$ as defined in Eq.~\eqref{eq:statesmissedapp}, \textcolor{black}{and using Eq.~\eqref{eq:expectations},}  we have
\begin{align}
    \ev{U_\kappa^2}{\Psi_{\text{all}}} &= \frac{1}{2}\pqty{3-2\sqrt{2}}\pqty{\kappa\Re(\alpha) - \frac{\Re(\beta)}{\kappa}}^2
    + \frac{1}{2}\pqty{\kappa^2 + \kappa^{-2}} + \frac{1}{2}\pqty{\cos\theta - \sin\theta}^2 \ev{n}_{\psi_\mathbf{n}}\\
    \ev{V_\kappa^2}{\Psi_{\text{all}}} &= \frac{1}{2}\pqty{3-2\sqrt{2}}\pqty{\kappa\Im(\alpha) + \frac{\Im(\beta)}{\kappa}}^2
    + \frac{1}{2}\pqty{\kappa^2 + \kappa^{-2}} + \frac{1}{2}\pqty{\cos\theta + \sin\theta}^2 \ev{n}_{\psi_\mathbf{n}}.
\end{align}
\end{widetext}
Therefore, 
\begin{equation}
\begin{aligned}
    \ev{U^2_\kappa} + \ev{V^2_\kappa} &= \frac{\kappa^2+\kappa^{-2}}{2}\pqty{\pqty{3-2\sqrt{2}}\sigma^2 + 2} + \ev{n}_{\psi_\mathbf{n}}\\
    &\geq \frac{\kappa^2+\kappa^{-2}}{2}\frac{\sigma^2}{1+\sigma^2},
    \end{aligned}
\end{equation}
where we have used that $\ev{n}_{\psi_\mathbf{n}} \geq 0$ and $\pqty{3-2\sqrt{2}}\sigma^2 + 2 \geq \frac{\sigma^2}{1+\sigma^2}$ for all $\sigma$. As such, none of these states will be found to be entangled by this criterion.

\subsection{Application to criterion by Hillery and Zubairy\textcolor{black}{, and Nha and Kim}\label{apd:SD-criteria}}
In the criterion by \citet{hillery2006entanglement}, a state is entangled if
\begin{equation}\label{eq:hillzubcrit_app}
    \abs{\ev{a_1a_2^\dag}}^2 > \ev{a_1^\dag a_1 a_2^\dag a_2}, 
\end{equation}

Using Eq.~\eqref{eq:expectations}, Hillery and Zubairy's criterion for these states reduces to the condition
\begin{equation}\label{eq:SD-condition}
\ev{(\Delta n)^2}_{\Psi_{\mathbf{n}}} < \ev{n}_{\Psi_{\mathbf{n}}},
\end{equation}
where $\ev{(\Delta n)^2}_{\Psi_{\mathbf{n}}} \equiv \ev{n^2}_{\Psi_{\mathbf{n}}}-\ev{n}_{\Psi_{\mathbf{n}}}^2$.

Meanwhile, defining the su(2) operators $J_x = (a_1^\dag a_2 + a_1 a_2^\dag)/2$ and $J_y = (a_1^\dag a_2 - a_1 a_2^\dag)/2i$, the entanglement condition by \citet{SU2entanglement} is
\begin{equation}
    \bqty{1+4\pqty{\Delta J_x}^2}\bqty{1+4\pqty{\Delta J_y}^2} < \pqty{1 + \ev{a_1^\dag a_1 + a_2^\dag a_2}}^2.
\end{equation}
Once again with Eq.~\eqref{eq:expectations}, their criterion simplifies to Eq.~\eqref{eq:SD-condition}.

Therefore, both criteria reduces to the same entanglement condition for the family of states we consider. For brevity, let us call these criteria the HZ/NK criteria. There are some states of the form Eq.~\eqref{eq:statesmissedapp} that satisfy Eq.~\eqref{eq:SD-condition}, which can therefore be detected by the HZ/NK criteria. We take a look at some specific examples of states in the $K=3$ case to study the applicability of the HZ/NK criteria in comparison to our DEW.

\subsubsection{Comparison to truncated maximally-violating states}
A natural family of states that can be detected by our DEW is for $\ket{\Phi_\mathbf{n}}$ where $\ket{\Psi_\mathbf{n}}$ is the maximally-violating eigenstate of $Q_3$ in the $a_+$ basis, truncated at $n_+ \leq 3\mathbf{n}$. For any $\mathbf{n}$, we can construct $Q_3$ explicitly as a matrix, and find $\ket{\Psi_\mathbf{n}}$ using standard eigenvalue solvers \cite{zaw2022detecting}.

\begin{figure}
    \centering
    \includegraphics[width=\columnwidth]{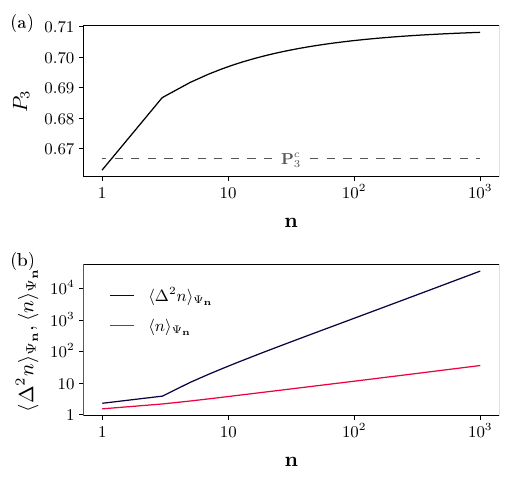}
    \caption{\label{fig:maxeig}(a) Observed violation of the truncated maximally-violating states. All states with $\mathbf{n} \geq 2$ violate the classical bound, and hence can be detected by our DEW for a certain range of $\theta$. (b) The standard deviation $\ev{(\Delta n)^2}_{\Psi_{\mathbf{n}}}$ and average $\ev{n}_{\Psi_{\mathbf{n}}}$ for these states. For all $\mathbf{n}$, $\ev{(\Delta n)^2}_{\Psi_{\mathbf{n}}} \geq \ev{n}_{\Psi_{\mathbf{n}}}$: none of these states satisfy Eq.~\eqref{eq:SD-condition}, so they will not be detected by the HZ/NK criteria.}
\end{figure}

The observed values of $P_3$ are plotted in Fig.~\ref{fig:maxeig}(a). All states with $\mathbf{n} \geq 2$ violate the classical bound, and hence can be detected by our DEW. The simplest state that can be detected is $\ket{\Phi_{\mathbf{n}=2}}$ where $\ket{\Psi_{\mathbf{n}=2}}_+ \propto \sqrt{16}\ket{0}_+ - \sqrt{21}\ket{3}_+ + \sqrt{5}\ket{6}_+$. On the other hand, we find that $\ev{(\Delta n)^2}_{\Psi_{\mathbf{n}}} \geq \ev{n}_{\Psi_{\mathbf{n}}}$ for all $\mathbf{n}$, as plotted in Fig.~\ref{fig:maxeig}(b). Therefore, none of these truncated maximally-violating states will be detected by the HZ/NK criteria.

\subsubsection{Comparison to entangled three-level cat entangled states}
Consider the state $\ket{\Phi_\mathbf{n}}$ where
\begin{equation}\label{eq:cat-qutrit-normal}
    \ket{\Psi_{\mathbf{n}}}_+ = \frac{3 e^{-2\abs{\alpha}^2}}{\sqrt{\mathcal{N}_{\Psi_\alpha}}}
    \sum_{n=0}^\infty \frac{\pqty{\sqrt{2}\alpha}^{3n}}{\sqrt{\pqty{3n}!}} \ket{3n}_+,
\end{equation}
with $\mathcal{N}_{\Psi_\alpha} = 3\bqty\big{
1 + 2{\cos}(\sqrt{3}\abs{\alpha}^2/2)e^{-3\abs{\alpha}^2/2}
}$ and $\alpha \in \mathcal{C}$ in general. When $\theta = \pi/4$, the state in the $\{a_1,a_2\}$ basis is
\begin{equation}\label{eq:cat-qutrit}
\begin{aligned}
    \ket{\Phi_{\mathbf{n}}} = \frac{1}{\sqrt{\mathcal{N}_{\Psi_\alpha}}}\bigg(
    &\lvert{\alpha e^{-i\frac{2\pi}{3}}}\rangle_1\otimes 
        \lvert{\alpha e^{-i\frac{2\pi}{3}}}\rangle_2 \\
    &{}+{}
        \lvert{\alpha}\rangle_1\otimes 
        \lvert{\alpha}\rangle_2 \\
    &{}+{}
        \lvert{\alpha e^{i\frac{2\pi}{3}}}\rangle_1\otimes 
        \lvert{\alpha e^{i\frac{2\pi}{3}}}\rangle_2
    \bigg).
\end{aligned}
\end{equation}
This is an entangled three-level cat state, with equally-separated coherent states in place of orthogonal qutrit states, analogous to entangled two-level cat states of the form $\propto \ket{\alpha}\otimes\ket{\alpha} + \ket{-\alpha}\otimes\ket{-\alpha}$ \cite{Two-Party-Cat-1,Two-Party-Cat-2}. Note that the state is entangled even when $0<\theta<\pi/4$, except that the terms in Eq.~\eqref{eq:cat-qutrit} would take the form $\lvert{\sqrt{2}\alpha\cos\theta e^{i\frac{2\pi k}{3}}}\rangle_1\otimes\lvert{\sqrt{2}\alpha \sin\theta e^{i\frac{2\pi k}{3}}}\rangle_2$ for $k \in \{-1,0,1\}$.

\begin{figure}
    \centering
    \includegraphics{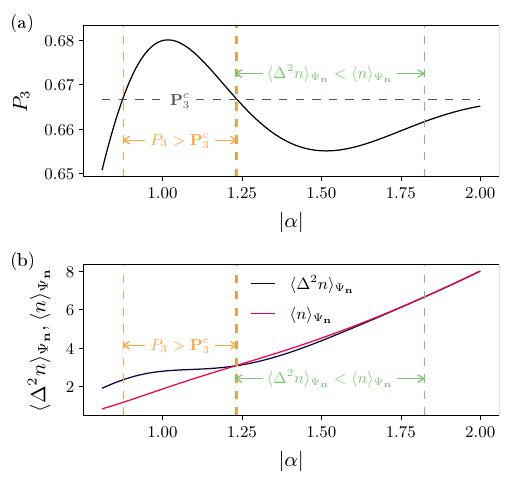}
    \caption{\label{fig:coherent-qutrit}(a) Observed violation of the three-level cat-like states as defined in Eq.~\eqref{eq:cat-qutrit-normal}, with $\alpha = -\abs{\alpha}$. States with $0.88 \lesssim \abs{\alpha} \lesssim 1.23$ violate the classical bound, and hence can be detected by our DEW for a certain range of $\theta$. (b) The standard deviation $\ev{(\Delta n)^2}_{\Psi_{\mathbf{n}}}$ and average $\ev{n}_{\psi_{\mathbf{n}}}$ for these states. For $1.23 \lesssim\abs{\alpha} \lesssim 1.82$, Eq.~\eqref{eq:SD-condition} is satisfied, so can be detected by Hillery and Zubairy's and Nha and Kim's criterion. Both criteria can be used at a small region around $\abs{\alpha} \sim 1.23$.}
\end{figure}

We plot the results for $\alpha = -\abs{\alpha}$ in Fig.~\ref{fig:coherent-qutrit}. These states can be detected by our DEW for $0.88 \leq \abs{\alpha} \leq 1.23$, and detected by the HZ/NK criteria for $1.23 \leq \abs{\alpha} \leq 1.82$, with a small region at $\abs{\alpha} \sim 1.23$ where it can be detected by both criteria. Therefore, one would choose the criteria to be used based on the magnitude of $\alpha$.

\section{Genuine multipartite entanglement}
Our DEW is naturally extendable to the multipartite scenario. We give an example in the tripartite case, with the physical modes $\{b_1,b_2,b_3\}$. The precession protocol is taken to be performed on the normal mode $b_+ = (b_1+b_2+b_3)/\sqrt{3}$.

For the state $\rho = \rho_{1}\otimes\rho_{23}$ separable over the $\{b_1|b_2,b_3\}$ partition, its Wigner function is $W_\rho(\beta_1,\beta_2,\beta_3) = W_{\rho_1}(\beta_1) W_{\rho_{23}}(\beta_2,\beta_3)$. Here, $\beta_j$ is the phase-space variable that corresponds to mode $b_j$. With a local transformation $\rho_{23} \to \rho_{23}'$ where $b_2 \to (b_2+b_3)/\sqrt{2}$ and $b_3 \to (b_2-b_3)/\sqrt{2}$, we have
\begin{equation}
W_\rho(\beta_1,\beta_2,\beta_3) = W_{\rho_1}(\beta_1) W_{\rho'_{23}}\pqty{\tfrac{\beta_2+\beta_3}{\sqrt{2}},\tfrac{\beta_2-\beta_3}{\sqrt{2}}}.
\end{equation}
Let us define the auxiliary modes $b_{-1} = (2b_1-b_2-b_3)/\sqrt{6}$ and $b_{-2} = (b_2-b_3)/\sqrt{2}$. Then, the Wigner function of $\rho_+ \equiv \tr_{-1,-2}(\rho)$, the reduced state of $\rho$ in the normal mode $b_+$, is found by integrating over $\beta_{-1}$ and $\beta_{-2}$, the phase space variables of the auxiliary modes:
\begin{align}
W_{\rho_+}(\beta_+) &= \int\dd{\beta_{-1}}\int\dd{\beta_{-2}}
W_\rho(\beta_+,\beta_{-1},\beta_{-2}) \nonumber\\
&= \int\dd{\beta_{-1}} W_{\rho_1}\pqty{\tfrac{1}{\sqrt{3}}\beta_+ + \sqrt{\tfrac{2}{3}}\beta_{-1}} \nonumber\\ &\qquad\times\underbrace{\int\dd{\beta_{-2}}W_{\rho'_{23}}\pqty{\sqrt{\tfrac{2}{3}}\beta_+ - \frac{1}{\sqrt{3}}\beta_{-1},\beta_{-2}}}_{\equiv W_{\rho''_{23}}\pqty{\sqrt{\tfrac{2}{3}}\beta_+ - \tfrac{1}{\sqrt{3}}\beta_{-1}}}\nonumber\\\label{eq:tri-is-bi}
&= \int\dd{\beta_{-1}} W_{\rho_1}\pqty{\tfrac{1}{\sqrt{3}}\beta_+ + \sqrt{\tfrac{2}{3}}\beta_{-1}}\\ &\qquad\qquad\qquad\times W_{\rho''_{23}}\pqty{\sqrt{\tfrac{2}{3}}\beta_+ - \tfrac{1}{\sqrt{3}}\beta_{-1}}.\nonumber
\end{align}
Here, $\rho''_{23} \equiv \tr_{-2}(\rho_{23}')$ is a local operation on $\rho_{23}'$. With the identification $a_+ \leftrightarrow b_+$ and $a_-\leftrightarrow b_{-1}$, we can recognize Eq.~\eqref{eq:tri-is-bi} as the original bipartite DEW with the mixing angle $\theta = \atan(1/\sqrt{2})$. Denoting $\mathbf{P}_K^{\text{SEP}}(\theta) \equiv \max_{\rho_{\text{SEP}}}\tr(Q_K\rho_{\text{SEP}})$, where $\rho_{\text{SEP}}$ is separable in the $\{a_1,a_2\}$ basis and $Q_K$ is defined in the mode $a_+$ for the original bipartite case, this implies that $P_K \leq \mathbf{P}_K^{\text{SEP}}(\theta=\atan(1/\sqrt{2}))$ for the precession protocol performed in the $b_+$ mode with the state $\rho_1 \otimes \rho_{23}$.

At the same time, we could have similarly defined $b'_{-1} = (-b_1+2b_2-b_3)/\sqrt{6}$ and $b'_{-2} = (b_1-b_3)/\sqrt{2}$, which tests for entanglement over the $\{b_2|b_3,b_1$\} partition, or $b''_{-1} = (-b_1-b_2+2b_3)/\sqrt{6}$ and $b''_{-2} = (b_1-b_2)/\sqrt{2}$, which tests for entanglement over the $\{b_3|b_1,b_2\}$ partition. That is, if the state is separable over \emph{any} partition $\{b_j|b_k,b_l\}$, then the quantum score $P_K$ is bounded above by $\mathbf{P}_K^{\text{SEP}}(\theta=\atan(1/\sqrt{2}))$.

Note that in all three cases, we are still performing the protocol on the normal mode $b_{+}$: we have merely grouped the auxiliary modes differently, which are traced over and does not affect the reduced state $\rho_+$. Hence, for all states of the form
\begin{equation}
\begin{aligned}
\rho &= \sum_j p^{(j)}_1 \rho^{(j)}_{1} \otimes \rho^{(j)}_{2,3}
+ \sum_k p^{(k)}_2 \rho^{(k)}_{2} \otimes \rho^{(k)}_{3,1}\\
&\qquad{}+{} \sum_l p^{(l)}_3 \rho^{(l)}_{3} \otimes \rho^{(l)}_{1,2},
\end{aligned}
\end{equation}
where $\sum_j p^{(j)}_1 + \sum_k p^{(k)}_2 + \sum_l p^{(l)}_3 = 1$, we have $P_K \leq \mathbf{P}_K^{\text{SEP}}(\theta=\atan(1/\sqrt{2}))$ when the precession protocol is performed on the $b_+$ mode. In other words, achieving the quantum score $P_K > \mathbf{P}_K^{\text{SEP}}(\theta=\atan(1/\sqrt{2}))$ certifies \emph{genuine tripartite entanglement} \cite{Genuine-Multipartite-Entanglement}.

While we do not have an analytical formula for $\mathbf{P}_K^{\text{SEP}}(\theta=\atan(1/\sqrt{2}))$, the methods laid out in Section~\ref{apd:SDP} can be used to find its value for any $K$ under some energy constraints. In particular, for $K=3$ with $n_+,n_- \leq \mathbf{n}=11$, we can read off Fig.~\ref*{fig:Heatmap of LogNeg} in the main text to find $\mathbf{P}_K^{\text{SEP}}(\theta=\atan(1/\sqrt{2})) = \mathbf{P}_3^c$.

This procedure can be further extended to multiple parties with steps similar to the above by considering all possible bipartitions. For those cases, the value of $\mathbf{P}_K^{\theta}$ at other mixing angles $\theta$ will be needed.

\end{document}